# Exact Model for Mode-Dependent Gains and Losses in Multimode Fiber

Keang-Po Ho, *Senior Member, IEEE*

*Abstract*—In the strong mode coupling regime, the model for mode-dependent gains and losses (collectively referred as MDL) of a multimode fiber is extended to the region with large MDL. The MDL is found to have the same statistical properties as the eigenvalues of the summation of two matrices. The first matrix is a random Gaussian matrix with standard deviation proportional to the accumulated MDL. The other matrix is a deterministic matrix with uniform eigenvalues proportional to the square of the accumulated MDL. The results are analytically correct for fibers with two or large number of modes, and also numerically verified for other cases.

## I. INTRODUCTION

CONVENTIONALLY for short-distance links [1]-[2], multimode fiber (MMF) may be used for long distance systems. Using mode-division multiplexing (MDM) in MMF [3]-[7], the channel capacity is ideally directly proportional to the number of modes.

The group delays of different modes in an MMF may have slight difference, giving modal dispersion [8]-[10]. In the strong mode coupling regime, the statistics of mode-dependent group delays is the same as the eigenvalues of a zero-trace Hermitian Gaussian random matrix, or zero-trace Gaussian unitary ensemble (GUE) [10].

Different modes in an MMF may potentially have different losses. The mode-dependent gains or losses (collectively referred to here as MDL) may also induce by optical components in the systems, especially optical amplifiers. MDL poses a fundamental limit to the system performance [11][12]. The extreme case of high MDL is equivalent to a reduction in the number of modes, leading to a proportional reduction in the channel capacity.

If the MMF comprises $K$ independent, statistically identical sections, each with uncoupled MDL variance $\sigma_g^2$, the accumulated MDL of the fiber link is defined as $\xi = \sqrt{K}\sigma_g$. In the strong coupling regime and for large number of sections $K$, the statistical properties of MDL depend solely on the accumulated MDL $\xi$ and number of modes. Depending on number of modes, for accumulated MDL $\xi$ less than 10 to 15 dB, the overall MDL in logarithmic scale has a distribution the same as the zero-trace GUE with overall MDL standard deviation (STD) [11]

$$\sigma_{mdl} = \xi\sqrt{1 + \tfrac{1}{12}\xi^2} \ . \qquad (1)$$

Here, the statistics of the overall MDL is extended to large MDL region.

As the main proposition, the overall MDL in logarithmic scale has the same statistical properties as the eigenvalues of the summation of two matrices

$$\xi\mathbf{G} + \kappa_D\xi^2\mathbf{F}, \qquad (2)$$

where $\mathbf{G}$ is a zero-trace GUE, $\mathbf{F}$ is the a deterministic uniform matrix, and $\kappa_D$ is a constant between 1/3 and 1/2, depending on number of modes $D$. The zero-trace GUE $\mathbf{G}$ has unity eigenvalue variance. The uniform matrix $\mathbf{F}$ has its eigenvalues deterministically and uniformly between $\pm 1$. As an example, in two dimension, $\mathbf{F} = \text{diag}[1, -1]$ or other Hermitian matrices having the same eigenvalues.

In later parts of this paper, the summation (2) is proved analytically with $\kappa_2 = 1/3$ for two-mode fiber, equivalently the polarization-dependent loss (PDL) of single-mode fiber. For MMF with large number of modes, equivalently free random variables [13][14], the summation (2) is also valid with $\kappa_\infty = 1/2$. For MMF with other number of modes, numerical simulation finds an empirical relationship $\kappa_D = \tfrac{1}{2}D/(1+D)$. The eigenvalue distribution of (2) is compared with numerical simulation and good match is found with large MDL of $\xi = 20$ dB.

The remainder of this paper is organized as follows: Section II presents the statistics for MMF with large MDL. Section III is the conclusion. The statistics of (2) is derived in Appendix A. Appendix B uses the theory of free random variables to find $\kappa_\infty = 1/2$ and prove (2).

## II. MMF WITH LARGE MDL

Manufacturing variations, bends, mechanical stresses, thermal gradients and other effects cause coupling between different MMF modes [8][9]. A long-haul MDM system is in the strong-coupling regime, in which the overall fiber length is far longer than a correlation length over which the local eigenmodes can be considered constant. Strong mode coupling reduces the amount of modal dispersion, minimizing the processing complexity of the receiver [10]. Likewise, strong mode coupling reduces the amount of MDL, improving system performance and channel capacity [11]. Strong mode-coupling also helps to achieve frequency diversity [15] to close the gap between average and outage channel capacity. In the strong coupling regime, a fiber can be modeled as a concatenation of many independent sections [10]-[12], [15] that can be described by random matrices.

*A. Random Matrix Model*

Random matrices were used to study polarization-mode dispersion (PMD) or PDL in single-mode fiber [16]-[18]. An MMF can be modeled as the products of $K$ random matrices [10]-[12][15] that represent independent MMF sections, each having length at least equal to the correlation length. The overall transfer matrix of an MMF is $\mathbf{M}^{(t)} = \mathbf{M}^{(K)} \cdots \mathbf{M}^{(2)} \mathbf{M}^{(1)}$. Followed the models of [11][15], for an MMF supporting $D$ propagating modes[1], the matrix for the $k$th section $\mathbf{M}^{(k)}$ is the product of three $D \times D$ matrices $\mathbf{M}^{(k)} = \mathbf{V}^{(k)} \mathbf{\Lambda}^{(k)} \mathbf{U}^{(k)*}$, $k = 1, \ldots, K$. Here, $\mathbf{U}^{(k)}$ and $\mathbf{V}^{(k)}$ are random unitary matrices representing modal coupling at the input and output of the section, respectively, and $\mathbf{\Lambda}^{(k)}$ is a diagonal matrix representing modal gains or losses of the uncoupled modes in the $k$th section. Just include MDL here and ignore the frequency dependence, $\mathbf{\Lambda}^{(k)} = \mathrm{diag}\left[ e^{\frac{1}{2}g_1^{(k)}}, e^{\frac{1}{2}g_2^{(k)}}, \ldots, e^{\frac{1}{2}g_D^{(k)}} \right]$, where the vector $\mathbf{g}^{(k)} = \left( g_1^{(k)}, g_2^{(k)}, \ldots, g_D^{(k)} \right)$ describes the uncoupled MDL of each section.

When the overall matrix $\mathbf{M}^{(t)}$ is decomposed into $D$ spatial channels $\mathbf{M}^{(t)} = \mathbf{V}^{(t)} \mathbf{\Lambda}^{(t)} \mathbf{U}^{(t)*}$, with $\mathbf{U}^{(t)}$ and $\mathbf{V}^{(t)}$ as input and output unitary beam-forming matrices, we are interested in the statistics of

$$\mathbf{\Lambda}^{(t)} = \mathrm{diag}\left[ e^{\frac{1}{2}g_1^{(t)}}, e^{\frac{1}{2}g_2^{(t)}}, \ldots, e^{\frac{1}{2}g_D^{(t)}} \right], \quad (3)$$

or the statistics of the vector $\mathbf{g}^{(t)} = \left( g_1^{(t)}, g_2^{(t)}, \ldots, g_D^{(t)} \right)$ that is the logarithms of the eigenvalues of $\mathbf{M}^{(t)} \mathbf{M}^{(t)*}$ and quantifies the overall MDL of the MMF.

In the small yet practical MDL regime, the MDL statistics depend only on the number of modes and on the accumulated MDL $\xi$ via (1) [11]. MDL has the statistics of zero-trace GUE.

Compared $\sigma_{\mathrm{mdl}} = \xi\sqrt{1 + \xi^2/12}$ (1) with $\xi \mathbf{G} + \kappa_D \xi^2 \mathbf{F}$ (2), the GUE $\mathbf{G}$ gives the linear term of $\xi$ (or 1 inside the square root) to the overall MDL. With $\kappa_\infty = 1/2$, the uniform matrix $\mathbf{F}$ gives the nonlinear factor ($\xi^2/12$ inside the square root). To certain extended, the results in [11] approximated the zero-trace uniform matrix $\mathbf{F}$ using zero-trace GUE.

From the summation (2), the distribution of MDL is dominated by the GUE for small accumulated MDL $\xi$ and is determined by the uniform matrix $\mathbf{F}$ for large MDL. The zero-trace GUE $\mathbf{G}$ also has $D$ peaks but those peaks become more uniform and climaxing with the increase of MDL.

---
[1]Throughout this paper, "modes" include both polarization and spatial degrees of freedom. For example, the two-mode case can describe the two polarization modes in single-mode fiber.

Table I: Comparison between simulation and empirical $\kappa_D$

| $D$ | Simulation | $\frac{1}{2}\frac{D}{D+1}$ |
|---|---|---|
| 2 | 0.3338 | 0.3333 |
| 3 | 0.3734 | 0.3750 |
| 4 | 0.4000 | 0.4000 |
| 5 | 0.4151 | 0.4167 |
| 6 | 0.4289 | 0.4286 |
| 7 | 0.4376 | 0.4375 |
| 8 | 0.4433 | 0.4444 |
| 10 | 0.4541 | 0.4545 |
| 16 | 0.4705 | 0.4706 |
| 512 | 0.4984 | 0.4990 |
| $\infty$ | --- | 0.5000 |

*B. Large MDL Regime*

In the large MDL region, we have the proposition that the MDL has the same distribution as the eigenvalues of the summation (2). In Appendix A, the probability density function (p.d.f.) for the eigenvalues of $\mathbf{G} + \beta\mathbf{F}$ is derived and shown in Table II of Appendix A for arbitrary factor $\beta$. Compared $\mathbf{G} + \beta\mathbf{F}$ with (2), the statistics of (2) is obtained with $\beta = \kappa_D \xi$ and scaled by $\xi$.

With the MDL given by the eigenvalues of (2), the overall MDL variance is the summation of the variance of the GUE and uniform matrix

$$\sigma_{\mathrm{mdl}}^2 = \xi^2 + \frac{1}{3}\frac{D-1}{D+1}\kappa_D^2 \xi^4. \quad (4)$$

For PDL in single-mode fiber, equivalently $D = 2$, the exact model [19] give $\sigma_{\mathrm{mdl}} = \xi\sqrt{1 + \xi^2/9}$ or $\kappa_2 = 1/3$. For very large matrix, the theory in [11][20] gives the dependence of (1) or $\kappa_\infty = 1/2$. For other cases, the values of $\kappa_D$ are between 1/3 and 1/2 and can be found by numerical simulation that gives the empirical values of $\kappa_D = \frac{1}{2}D/(1+D)$. Table I compares the numerical with the theoretical/empirical values of $\kappa_D$. The numerical values are obtained by the mean of $\xi^{-1}\sqrt{(\sigma_{\mathrm{mdl}}/\xi)^2 - 1}$ for $\xi$ from 1 to 20 dB, the simulation is the same as that in [11]. Table I shows that the theoretical/empirical values $\kappa_D = \frac{1}{2}D/(1+D)$ are very accurate. In later part of this paper, only $\kappa_D = \frac{1}{2}D/(1+D)$ are used.

Using $\kappa_D = \frac{1}{2}D/(1+D)$ in (4), the overall MDL STD becomes

$$\sigma_{\mathrm{mdl}} = \xi\sqrt{1 + \frac{\xi^2}{12(1-D^{-2})}}. \quad (5)$$

The overall MDL STD (5) approaches (1) rapidly with the increase of dimension $D$, the same as the results in [11]. Using the theory of free random variables [13][14], the central limit for the products of large random matrix have a moment generating function given by (B.1) of Appendix B. In Appendix B, for large random matrix, the moment generating function (B.1) is found to have the same moments as the summation (2) with $\kappa_\infty = 1/2$.

Fig. 1 shows the exact theoretical p.d.f. from Table II (after scaling) as compared with the simulated p.d.f. for accumulated MDL of $\xi = 20$ dB. The simulation is the product of random matrices and is conducted similar to that in [11]. There are significant observable differences between the simulated results with the approximation in [11] for $\xi = 20$ dB.

Fig. 2 compares the simulated p.d.f. for $D = 16$, 64, and 512 with the theoretical p.d.f. for large matrix given by the moment generating function of (B.1). The moment generating function (B.1) is first converted to characteristic function. Inverse Fourier transform of the characteristic function numerically gives the theoretical p.d.f. of Fig. 2. Fig. 2 also shows the semicircle distribution with overall MDL given by (1).

Fig. 1 shows no difference between the p.d.f. of Table II and numerical simulation, providing the necessary validation that the eigenvalues of the summation (2) give the MDL distribution, even in the region with large MDL. Appendix B verifies that the summation (2) has the same statistics as the central limit of the product of free random variables, equivalently the products of large random matrices. Fig. 2 shows that the MDL for fiber with large number of modes has the same statistics as (B.1), equivalently the summation (2).

*C. Large PDL Regime*

For single-mode fiber that supports two polarization modes, the MDL for $D = 2$ is equivalent to PDL for single-mode fiber. Simulation in Fig. 1 matches with the theory in Table II very well. For PDL, the statistical properties of the followings are the same:

a. PDL for single-mode fiber with strong random mode coupling and accumulated PDL as $\xi$.
b. The eigenvalues of the summation of two two-dimensional matrices: a zero-trace GUE with eigenvalue STD of $\xi$ and a diagonal matrix with $\pm \xi^2/3$ as the diagonal elements.
c. The concatenation of a random Maxwellian distributed PMD with root-mean-square differential group delay (DGD) of $\xi$ and deterministic PMD with DGD of $\xi^2/3$.

According to Eq. (14) of [19], using the notation here, the exact PDL distribution is[2]

$$p_2^{(a)}(x) = \sqrt{\frac{6}{\pi}} \frac{x \sinh x}{\xi^3} \exp\left(-\frac{3x^2}{2\xi^2} - \frac{\xi^2}{6}\right), \quad x \geq 0, \quad (6)$$

Using Table II with $D = 2$, the p.d.f. $p_2(x)$ can be simplified to

$$p_2(x) = \sqrt{\frac{6}{\pi}} \exp\left[-\tfrac{3}{2}(x^2 + \beta^2)\right] \frac{x \sinh(3\beta x)}{4\beta}.$$

With $\beta = \xi/3$ and scale by $\xi$, the p.d.f. for the eigenvalues of (2) with $D = 2$ is

$$p_2^{(b)}(x) = \frac{1}{2}\sqrt{\frac{6}{\pi}} \frac{x \sinh x}{\xi^3} \exp\left(-\frac{3x^2}{2\xi^2} - \frac{\xi^2}{6}\right),$$

---
[2] The substitutions are $a = \gamma x$ and $t = \xi^2/3$.

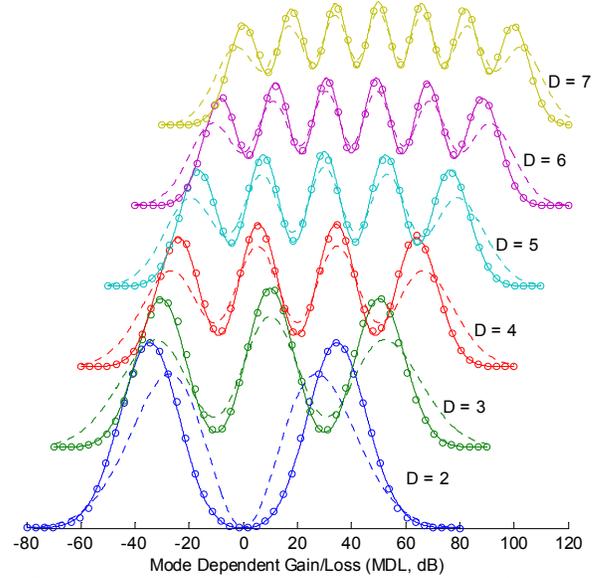

Fig. 1 The exact p.d.f. (solid curves) as compared with simulated (markers) and approximated (dotted curves, from [11]) p.d.f. for $\xi = 20$ dB.

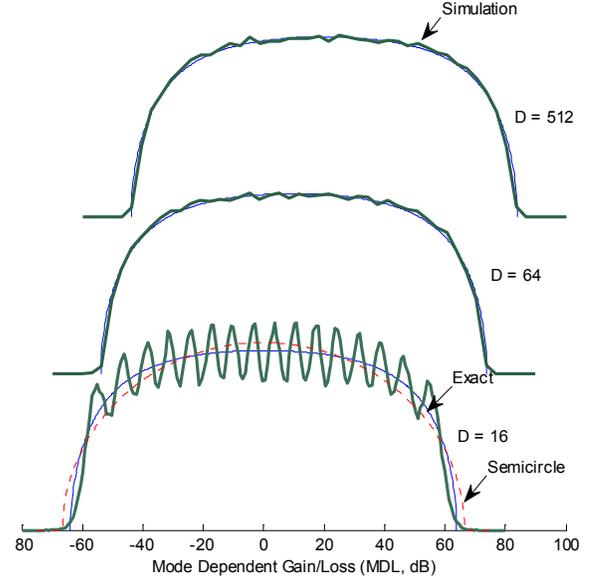

Fig. 2 The simulated p.d.f. for $D = 16$, 64, and 512 compared with the large matrix p.d.f. given by the moment generating function of (B.1). Semicircle distribution is also shown for comparison.

Note that $p_2^{(b)}(x)$ is 1/2 of $p_2^{(a)}(x)$ (6) because the p.d.f. of Table II is for both positive and negative $x$ instead of the one-sided p.d.f. in [19] for positive $x$ only. Here, we prove that the eigenvalues of (2) with $\kappa_2 = 1/3$ give the same p.d.f. as the exact PDL of [19].

Concatenation of random single-mode fiber with deterministic DGD has p.d.f. given by Eqs. (26) and (27) of [16]. Compared Eq. (26) of [16] with $p_2(x)$ in Table II, they

K.-P. Ho, Exact Model for MDL in MMF     4are very similar except scale factors. The p.d.f. of Eq. (27) of [16] is the same as $p_2^{(a)}(x)$ (6) using the notations here[3].

The p.d.f. (6) is a non-central chi distribution with three degrees of freedom. Maxwellian distribution is the "central" chi distribution with three degrees of freedom. The chi distribution is not as well-known as the chi-square distribution [21]. In $D = 2$, PDL has a non-central Maxwellian distribution but with very specific noncentrality parameter of $\xi/3$.

If the modal dispersion model of [10] is used, the summation (2) represents the concatenation of two fiber pieces: the first piece as a standard MMF fiber with sufficiently strong mode coupling and the second piece with deterministic group delay uniformly between $\pm \kappa_D \xi^2$. PMD is only a special case of modal dispersion in two-mode fiber. The group delays for modal dispersion are in certain normalized unit, corresponding to MDL, instead of time.

## III. CONCLUSION

To generalize from PDL of single-mode fiber, we find that the followings have the same statistics:

a. MDL in logarithmic scale for $D$-mode MMF in strong mode coupling regime with accumulated MDL of $\xi$.
b. The eigenvalues of the summation of two $D$-dimensional matrices: a zero-trace GUE with eigenvalue STD of $\xi$ and a diagonal matrix with $D$ diagonal elements uniformly between $\pm \kappa_D \xi^2$, where $\kappa_D = \tfrac{1}{2} D/(1+D)$.
c. The modal dispersion of the concatenation of two MMF pieces: one with strong mode coupling and group delay STD of $\xi$ and another with deterministic modal dispersion having group delay uniformly between $\pm \kappa_D \xi^2$.
d. The central limit for the product of $D \times D$ random matrices with accumulated variance in logarithmic scale of $\xi^2$.

The above statistics for MDL is derived analytically for two and many mode fibers, and verified numerically for other MMF with other number of modes.

## ACKNOWLEDGEMENT

For free random variables, the logarithmic of the products of free random variables is the free convolution of GUE (or semicircle distribution) and uniformly distributed random matrices given by (B.8) or (2) was first conjectured by Christian Sattlecker, TU Graz, Austria and suggested to the author by private communications. His later proof has substantial difference with that in Appendix B.

## APPENDIX A: STATISTICS OF $\mathbf{G} + \beta \mathbf{F}$

The eigenvalue statistics of (2) is the MDL distribution in logarithmic scale. When two independent conventional random variables sum together, the p.d.f. is the convolution of the two respective p.d.f. When the dimension of the matrix approaches infinity, the eigenvalue statistics of the sum of independent random matrices is call free convolution. In general, the eigenvalue statistics of (2) is the finite free convolution of two $D \times D$ random matrices.

The second uniform matrix $\mathbf{F}$ in (2) is in general also a random matrix with deterministic uniformly distributed eigenvalues of $\alpha_i = -1 + 2(i-1)/(D-1)$, $i = 1, \ldots, D$, and $\mathbf{F} = \mathbf{U}_F \,\mathrm{diag}\{-1, -1 + 2/(D-1), \ldots, +1\}\mathbf{U}_F^*$, where $\mathbf{U}_F$ is a random unitary matrix having the random eigenvectors of $\mathbf{F}$. The list of $\alpha_i$ depends on the dimension $D$. In practice, the random unitary matrix $\mathbf{U}_F$ may be absorbed to the corresponding random unitary matrix of the first Gaussian random matrix $\mathbf{G}$. To further simplify the derivation, the statistic of $\mathbf{X}$ in

$$\mathbf{U}\mathbf{X}\mathbf{U}^* = \mathbf{A} = \mathbf{G} + \beta \mathbf{F} \tag{A.1}$$

is derived for $\mathbf{G}$ as a Gaussian random matrix with unity eigenvalue variance and $\mathbf{F} = \mathrm{diag}\{-1, -1 + 2/(D-1), \ldots, +1\} = \mathrm{diag}\{\alpha_1, \alpha_2, \ldots, \alpha_D\}$ is a deterministic uniform diagonal matrix. In (A.1), $\mathbf{U}$ is a random unitary matrix and $\mathbf{X} = \mathrm{diag}\{x_1, x_2, \ldots x_D\}$ is an diagonal matrix with the eigenvalues of $\mathbf{A} = \mathbf{G} + \beta \mathbf{F}$. In the theory of this section, after certain normalization, the eigenvalues $x_1, x_2, \ldots, x_D$ correspond to the MDL of the MMF in logarithmic scale.

Form [22], the derivation of matrix elements is

$$d\mathbf{A} = |\Delta \mathbf{x}|^2 \, d\mathbf{X} d\mathbf{U} \tag{A.2}$$

where

$$\Delta \mathbf{x} = \prod_{i > j} (x_i - x_j). \tag{A.3}$$

The distribution of the elements for $\mathbf{G}$ is proportional to $\exp\!\left(-\tfrac{1}{2}\,\mathrm{tr}\,\mathbf{G}^2/\sigma_D^2\right)$ with $\sigma_D^{-2} = D - D^{-1}$ such that the eigenvalues of $\mathbf{G}$ has unity variance [10]. From (A.2), the distribution of $\mathbf{X}$ requires integration over $\mathbf{U}$. The integration is [22]

$$\int \exp\!\left[-\frac{\mathrm{tr}\!\left(\mathbf{U}\mathbf{X}\mathbf{U}^* - \beta\mathbf{F}\right)^2}{2\sigma_D^2}\right] d\mathbf{U} \propto \frac{1}{|\Delta \mathbf{x}|} \det\!\left[\exp\!\left(-\frac{(x_i - \beta\alpha_j)^2}{2\sigma_D^2}\right)\right]. \tag{A.4}$$

In the original form of (A.4) in [22], the integration is also proportional to $\prod_{i>j}(\alpha_i - \alpha_j)^{-1}$ that is a constant by itself. In (A.4), det[ ] is an determinant with $i, j$ elements given by a function.

Combining (A.2) and (A.4), the p.d.f. of the eigenvalues $x_1$, $x_2$, …, $x_D$ is

$$p_\mathbf{X}(\mathbf{x}) \propto |\Delta \mathbf{x}| \det\!\left[\exp\!\left(-\frac{(x_i - \beta\alpha_j)^2}{2\sigma_D^2}\right)\right]. \tag{A.5}$$

Both the ordered and unordered eigenvalues has the same distribution (A.5) but different normalization factor. The same as that in [10], the zero-trace constraint requires

---

[3] Using the room-mean-square value of the DGD, the average DGD is given by $2\sqrt{2/3\pi}\,\xi$. The deterministic DGD here is $|\tau_{\mathrm{det}}| = \xi^2/3$.

Table II Eigenvalue Distribution of $\mathbf{G} + \beta\mathbf{F}$

General Formula: $p_D(x) = \sum_{j=1}^{D} f_j^{(D)}(x,\beta) \exp\left[-\frac{D+1}{2}\left(x+\beta - 2\beta\frac{j-1}{D-1}\right)^2\right]$

| $D$ | $f_j^{(D)}(x,\beta)$ |
|---|---|
| 2 | $f_1^{(2)}(x,\beta) = -\sqrt{\frac{6}{\pi}}\frac{x}{4\beta}$, $f_2^{(2)}(x,\beta) = f_1^{(2)}(-x,\beta)$. |
| 3 | $f_1^{(3)}(x,\beta) = \sqrt{\frac{2}{\pi}}\frac{1}{8\beta^2}\left(3x^2 - \frac{3}{4} - \frac{1}{3}\beta^2\right)$, $f_2^{(3)}(x,\beta) = \sqrt{\frac{2}{\pi}}\frac{1}{\beta^2}\left(-\frac{3}{4}x^2 + \frac{3}{16} + \frac{1}{3}\beta^2\right)$, $f_3^{(3)}(x,\beta) = f_1^{(3)}(-x,\beta)$. |
| 4 | $f_1^{(4)}(x,\beta) = \sqrt{\frac{10}{\pi}}\frac{x}{2\beta^3}\left(\frac{1}{3}x^2 - \frac{1}{5} - \frac{1}{12}\beta^2\right)$, $f_2^{(4)}(x,\beta) = \sqrt{\frac{10}{\pi}}\frac{1}{2\beta^3}\left(-x^3 + \frac{3}{5}x + \frac{7}{12}x\beta^2 + \frac{5}{54}\beta^3\right)$, $f_3^{(4)}(x,\beta) = f_2^{(4)}(-x,\beta)$, $f_4^{(4)}(x,\beta) = f_1^{(4)}(-x,\beta)$. |
| 5 | $f_1^{(5)}(x,\beta) = \sqrt{\frac{3}{\pi}}\frac{1}{64\beta^4}\left(\frac{125}{6}x^4 - \frac{125}{6}x^2 - \frac{25}{3}\beta^2 x^2 + \frac{125}{72} + \frac{25}{18}\beta^2 + \frac{3}{10}\beta^4\right)$, $f_2^{(5)}(x,\beta) = \sqrt{\frac{3}{\pi}}\frac{1}{64\beta^4}\left(-\frac{250}{3}x^4 + \frac{250}{3}x^2 + \frac{175}{3}\beta^2 x^2 + 10\beta^3 x - \frac{125}{18} - \frac{175}{18}\beta^2 - \frac{63}{40}\beta^4\right)$, $f_3^{(5)}(x,\beta) = \sqrt{\frac{3}{\pi}}\frac{1}{\beta^4}\left(\frac{125}{64}x^4 - \frac{125}{64}x^2 - \frac{25}{16}\beta^2 x^2 + \frac{125}{768} + \frac{25}{96}\beta^2 + \frac{1}{5}\beta^4\right)$, $f_4^{(5)}(x,\beta) = f_2^{(5)}(-x,\beta)$, $f_5^{(5)}(x,\beta) = f_1^{(5)}(-x,\beta)$. |
| 6 | $f_1^{(6)}(x,\beta) = \sqrt{\frac{14}{\pi}}\frac{x}{8\beta^5}\left(\frac{27}{20}x^4 - \frac{27}{14}x^2 - \frac{3}{4}\beta^2 x^2 + \frac{81}{196} + \frac{9}{28}\beta^2 + \frac{1}{15}\beta^4\right)$, $f_2^{(6)}(x,\beta) = \sqrt{\frac{14}{\pi}}\frac{1}{2\beta^5}\left(-\frac{27}{16}x^5 + \frac{135}{56}x^3 + \frac{111}{80}\beta^2 x^3 + \frac{21}{100}\beta^3 x^2 - \frac{405}{784}x - \frac{333}{560}\beta^2 x - \frac{173}{1500}\beta^4 x - \frac{3}{100}\beta^3 - \frac{77}{9375}\beta^5\right)$, $f_3^{(6)}(x,\beta) = \sqrt{\frac{14}{\pi}}\frac{1}{\beta^5}\left(\frac{27}{16}x^5 - \frac{135}{56}x^3 - \frac{129}{80}\beta^2 x^3 - \frac{21}{100}\beta^3 x^2 + \frac{405}{784}x + \frac{387}{560}\beta^2 x + \frac{229}{750}\beta^4 x + \frac{3}{200}\beta^3 + \frac{364}{9375}\beta^5\right)$, $f_4^{(6)}(x,\beta) = f_3^{(6)}(-x,\beta)$, $f_5^{(6)}(x,\beta) = f_2^{(6)}(-x,\beta)$, $f_6^{(6)}(x,\beta) = f_1^{(6)}(-x,\beta)$. |
| 7 | $f_1^{(7)}(x,\beta) = \frac{1}{512\sqrt{\pi}\beta^6}\left(\frac{16807}{45}x^6 - \frac{16807}{24}x^4 - \frac{2401}{9}\beta^2 x^4 + \frac{16807}{64}x^2 - \frac{2401}{12}\beta^2 x^2 + \frac{1813}{45}\beta^4 x^2 - \frac{16807}{1536} - \frac{2401}{192}\beta^2 - \frac{1813}{360}\beta^4 - \frac{5}{7}\beta^6\right)$, $f_2^{(7)}(x,\beta) = \frac{1}{8\sqrt{\pi}\beta^6}\left(-\frac{16807}{480}x^6 + \frac{16807}{256}x^4 + \frac{2401}{72}\beta^2 x^4 + \frac{343}{81}\beta^3 x^3 - \frac{50421}{2048}x^2 - \frac{2401}{96}\beta^2 x^2 - \frac{686}{135}\beta^4 x^2 - \frac{343}{216}\beta^3 x - \frac{112}{243}\beta^5 x + \frac{16807}{16384} + \frac{2401}{1536}\beta^2 + \frac{343}{540}\beta^4 + \frac{1280}{15309}\beta^6\right)$, $f_3^{(7)}(x,\beta) = \frac{1}{4\sqrt{\pi}\beta^6}\left(\frac{16807}{384}x^6 - \frac{84035}{1024}x^4 - \frac{55223}{1152}\beta^2 x^4 - \frac{343}{81}\beta^3 x^3 + \frac{252105}{8192}x^2 + \frac{55223}{1536}\beta^2 x^2 + \frac{40621}{3456}\beta^4 x^2 + \frac{343}{216}\beta^3 x + \frac{469}{243}\beta^5 x - \frac{84035}{65536} - \frac{55223}{24576}\beta^2 - \frac{40621}{27648}\beta^4 - \frac{230945}{1959552}\beta^6\right)$, $f_4^{(7)}(x,\beta) = \frac{1}{\sqrt{\pi}\beta^6}\left(-\frac{16807}{1152}x^6 + \frac{84035}{3072}x^4 + \frac{2401}{144}\beta^2 x^4 - \frac{84035}{8192}x^2 - \frac{2401}{192}\beta^2 x^2 - \frac{343}{72}\beta^4 x^2 + \frac{84035}{196608} + \frac{2401}{3072}\beta^2 + \frac{343}{576}\beta^4 + \frac{2}{7}\beta^6\right)$, $f_5^{(7)}(x,\beta) = f_3^{(7)}(-x,\beta)$, $f_6^{(7)}(x,\beta) = f_2^{(7)}(-x,\beta)$, $f_7^{(7)}(x,\beta) = f_1^{(7)}(-x,\beta)$. |

$$x_1 + x_2 + \ldots + x_D = 0. \quad (A.6)$$

Disregard the order of the eigenvalues, with $x_D$ determined by (A.6), the distribution of MDL is given by

$$p_D(x_1) \propto \int \cdots \int |\Delta\mathbf{x}| \det\left[\exp\left(-\frac{(x_i - \beta\alpha_j)^2}{2\sigma_D^2}\right)\right] dx_2 \cdots dx_{D-1}. \quad (A.7)$$

In the straightforward calculation, the determinant in (A.7) seems to have $D!$ terms. Due to symmetric nature of each variable, all terms with the same factor of $x_1 - \beta\alpha_j$ is identical. As shown in Table II, the general form of the distribution (A.7) is

$$p_D(x) = \sum_{j=1}^{D} f_j^{(D)}(x,\beta) \exp\left[-\frac{D+1}{2}(x - \beta\alpha_j)^2\right]. \quad (A.8)$$

where the factor $(D+1)/2$ is given by [11] when $\beta = 0$ and $f_j^{(D)}(x,\beta)$ is a polynomial depending on $x$ and $\beta$. Because of the symmetric nature of the variable, we have

$$f_j^{(D)}(x,\beta) = f_{D-j+1}^{(D)}(-x,\beta). \quad (A.9)$$

In the calculation, as an example, for $j = 1$,

$$f_1^{(D)}(x_1, \beta) \exp\left[-\frac{D+1}{2}(x_1+\beta)^2\right]$$
$$\propto \int \cdots \int |\Delta \mathbf{x}| \prod_{i=1}^{D} \exp\left(-\frac{(x_i - \beta\alpha_i)^2}{2\sigma_D^2}\right) \prod_{i=2}^{D-1} \mathrm{d}x_i \quad (A.10)$$

All expression in the form of (A.10) can be found analytically in the method similar to that in [10] and presented in Table II. The p.d.f. $p_D(x)$ is normalized by integrated to unity.

## APPENDIX B: FREE RANDOM VARIABLE $\xi \mathbf{G} + \frac{1}{2}\xi^2 \mathbf{F}$

In [20], the product of large random matrices has a central limit. The central limit is derived based on the theory of free random variables [13][14] that is asymptotically equivalent to the eigenvalue statistics of large random matrices. In logarithmic scale and using the notation here, the central limit has moment generating function of [20]

$$M_{\log \mathbf{Y}}(s) = \exp\left(\tfrac{1}{2}\xi^2 s\right)_1F_1(1-s; 2; -\xi^2 s), \quad (B.1)$$

where $_1F_1(a,b;z)$ is the confluent hypergeometric function and $\mathbf{Y}$ is the central limit of the product of free random variables. Here, the $n$th moment of (B.1) is derived and find to be the same as the $n$th moment of (2) with $\kappa_\infty = 1/2$.

The confluent hypergeometric function has the Buchholz expansion [23]

$$_1F_1(1-a, 2; z) = 2\exp\left(-\frac{z}{2}\right)\sum_{k=0}^{\infty} p_k(z) \frac{J_{k+1}(2\sqrt{az})}{(2\sqrt{az})^{k+1}} \quad (B.2)$$

where $J_k(z)$ is the Bessel function of the first kind and $p_k(z)$ is the Buchholz polynomial given by [24][25]

$$\exp\left[-\tfrac{1}{2}z\left(\coth t - \tfrac{1}{t}\right)\right] = \sum_{k=0}^{\infty} p_k(z)\left(-\tfrac{t}{z}\right)^k \quad (B.3)$$

or

$$p_k(z) = \frac{(-z)^k}{k!} \frac{\mathrm{d}^k}{\mathrm{d}t^k}\exp\left(-\tfrac{1}{2}zH_v(t)\right)\Big|_{t=0} \quad (B.4)$$

with $H_v(t) = \coth t - 1/t$. Using Buchholz polynomial given by (B.4), we obtain

$$M_{\log \mathbf{Y}}(s) = 2\sum_{k=0}^{\infty} \frac{\xi^{2k} s^k}{k!} \frac{\mathrm{d}^k}{\mathrm{d}t^k}\exp\left(\tfrac{1}{2}\xi^2 s H_v(t)\right)\Big|_{t=0} E_{k+1}(2\xi s) \quad (B.5)$$

where

$$E_k(x) = \frac{I_k(x)}{x^k} = \frac{1}{2^k}\sum_{m=0}^{\infty}\frac{1}{m!(m+k)!}\left(\frac{x}{2}\right)^{2m} \quad (B.6)$$

with $I_k(z)$ as the modified Bessel function of the first kind.

The $n$th moment of $\log \mathbf{Y}$ is given by the $n$th derivative of $M_{\log \mathbf{Y}}(s)$ (B.5). Alternatively, the $n$th moment of $\log \mathbf{Y}$ is $n!$ of the coefficient of $s^n$ when $M_{\log \mathbf{Y}}(s)$ is expanded in a power series of $s$. In each term of (B.5) with $k$ less than $n$, $s^k$ is given in each term, $s^i$ with $i$ less than $n-k$ may be given by

$$\frac{1}{i!}\frac{\mathrm{d}^i}{\mathrm{d}s^i}\frac{\mathrm{d}^k}{\mathrm{d}t^k}\exp\left(\tfrac{1}{2}\xi^2 s H_v(t)\right)\Big|_{t=s=0} = \frac{\xi^{2i}}{2^i i!}\frac{\mathrm{d}^k}{\mathrm{d}t^k}H_v^i(t)\Big|_{t=0}$$

and $s^{n-k-i}$ is provided by $E_{k+1}(2\xi s)$. The three terms combine to give the coefficient of $s^n$. The $n$th moment of $\log \mathbf{Y}$ is

$$2n! \sum_{k=0}^{n} \frac{\xi^{2k}}{k!} \sum_{i=0}^{n-k} \frac{\xi^{2i}}{2^i i!}\frac{\mathrm{d}^k}{\mathrm{d}t^k}H_v^i(t)\Big|_{t=0}$$
$$\times \frac{\xi^{n-k-i}}{2^{k+1}[(n-k-i)/2]![(n+k-i)/2+1]!}$$

or

$$n!\sum_{k=0}^{n}\frac{1}{k!}\sum_{i=0}^{n-k}\frac{\mathrm{d}^k}{\mathrm{d}t^k}H_v^i(t)\Big|_{t=0} \frac{\xi^{n+k+i}}{i!2^{k+i}[(n-k-i)/2]![(n+k-i)/2+1]!} \quad (B.7)$$

The $n$th moment is derived for the free random variable given by

$$\mathbf{S} = \xi \mathbf{G} + \tfrac{1}{2}\xi^2 \mathbf{F}. \quad (B.8)$$

The statistics of the summation of free random variables can be found by the $R$-transform [13][14]. The Cauchy-Stieltjes transform of a measure is given by the expectation

$$G(z) = E\left\{\frac{1}{z-X}\right\} = \sum_{k=0}^{\infty}\frac{m_k}{z^{k+1}} \quad (B.9)$$

where $X$ is random variable with moments of $m_k = E\{X^k\}$, and $E\{\}$ denotes expectation. In conventional random variable, the Cauchy-Stieltjes transform (B.9) is well-defined. In free random variable, the transform (B.9) is with respect to the eigenvalue statistics. The $R$-transform is defined by the algebra relationship [13][14]

$$G\left(R(z) + \frac{1}{z}\right) = z. \quad (B.10)$$

The $R$-transform for the free convolution, like (B.8), is the sum of the $R$-transform for each individual component. $\mathbf{G}$ has a semicircle distribution with radius of 2. The $R$-transform for $\xi \mathbf{G}$ is $\xi^2 z$. The Cauchy-Stieltjes transform for $\tfrac{1}{2}\xi^2 \mathbf{F}$ is $2\xi^{-2}\coth^{-1}(2z/\xi^2)$. The $R$-transform for $\tfrac{1}{2}\xi^2 \mathbf{F}$ is $\tfrac{1}{2}\xi^2 \coth\left(\tfrac{1}{2}\xi^2 z\right) - 1/z$. The $R$-transform for $\mathbf{S}$, given by (B.8), is $\xi^2 z + \tfrac{1}{2}\xi^2 \coth\left(\tfrac{1}{2}\xi^2 z\right) - 1/z$. If $G_S(z)$ is the Cauchy-Stieltjes transform for $\mathbf{S}$, using (B.10), the algebraic inverse for $G_S(1/z)$ is given by $\left[\xi^2 w + \tfrac{1}{2}\xi^2 \coth\left(\tfrac{1}{2}\xi^2 w\right)\right]^{-1}$.

Compared with (B.9), the $(n+1)$th term of $G_S(1/z)$ is the $n$th moment of $\mathbf{S}$. If the inverse of $G_S(1/z)$ is expressed as $w/\phi(w)$, using the Lagrange inversion theorem [26], the $(n+1)$th term of $G_S(1/z)$ is given by

$$\frac{1}{(n+1)!}\frac{\mathrm{d}^n}{\mathrm{d}w^n}\phi(w)^{n+1}\Big|_{w=0}$$

with

$$\phi(w) = \xi^2 w^2 + 1 + \tfrac{1}{2}\xi^2 w H_v\left(\tfrac{1}{2}\xi^2 w\right), \quad (B.11)$$

where $H_v$ is defined in (B.4). The $n$th moment of $\mathbf{S}$ is just $1/(n+1)$ of the $n$th term in

$$\phi^{n+1}(w) = \sum_{k=0}^{n+1}\binom{n+1}{k}\left[\tfrac{1}{2}\xi^2 w H_v\left(\tfrac{1}{2}\xi^2 w\right)\right]^k \left(1+\xi^2 w^2\right)^{n+1-k}. \quad (B.12)$$

For the coefficient for $w^n$, the first term in (B.12) gives $w^k$ for $k = 0$ to $n$, the second term provides

$$\left.\frac{\xi^{2i}}{i!2^i}\frac{d^i}{dw^i}H_v^k(w)\right|_{w=0} w^i$$

and the third term of (B.12) gives $w^{n-k-i}$. The $n$th moment of **S** becomes

$$\frac{1}{n+1}\sum_{k=0}^{n}\binom{n+1}{k}\frac{\xi^{2k}}{2^k}\sum_{i=0}^{n-k}\left.\frac{\xi^{2i}}{i!2^i}\frac{d^i}{dw^i}H_v^k(w)\right|_{w=0} \\ \times \binom{n+1-k}{(n-k-i)/2}\xi^{n-k-i} \quad (B.13)$$

or

$$\frac{1}{n+1}\sum_{k=0}^{n}\binom{n+1}{k}\sum_{i=0}^{n-k}\left.\frac{\xi^{n+k+i}}{i!2^{k+i}}\frac{d^i}{dw^i}H_v^k(w)\right|_{w=0}\binom{n+1-k}{(n-k-i)/2}. \quad (B.14)$$

Compared (B.14) with (B.7), (B.14) is the same as (B.7) by swap the index of $i$ and $k$, exchange the order of summation, and using the relationship of

$$\frac{1}{n+1}\binom{n+1}{k}\binom{n+1-k}{(n-k-i)/2} \\ = \frac{n!}{k!}\times\frac{1}{[(n-k-i)/2]![(n+k-i)/2+1]!}.$$

With the equivalent of (B.7) and (B.14), as the moments uniquely determine a distribution, we may conclude that the product of random matrices, in logarithmic scale, is the free summation of a Gaussian unitary ensemble and a uniformly distributed random matrix in the form of (B.8).